\documentclass{ws-ijmpd}

\begin{document}

\markboth{Vijay Singh and Aroonkumar Beesham}
{LRS Bianchi I model with perfect fluid equation of state}

\catchline{}{}{}{}{}
%
%%%%%%%%%%%%%%%%%%%%%%%%%%%%%%%%%%%%%%%%%%%%%%%%%%%%%%%%%%%%%%%%%%%%

\title{LRS BIANCHI I MODEL WITH PERFECT FLUID\\EQUATION OF STATE}
%\footnote{For the title, try not to use more than 3 lines. Typeset the title in 10~pt Times roman, uppercase and boldface.}  }

\author{VIJAY SINGH}
%\footnote{Typeset names in 8~pt roman, uppercase. Use the footnote to indicate the present or permanent address of the author.}}

\address{Department of Mathematical Sciences,\\
University of Zululand, Private Bag X1001\\
Kwa-Dlangezwa, 3886, South Africa.
%\footnote{State completely without abbreviations, the affiliation and mailing address, including country. Typeset in 8~pt Times italic.}
\\
gtrcosmo@gmail.com}

\author{AROONKUMAR BEESHAM}

\address{Department of Mathematical Sciences,\\
University of Zululand, Private Bag X1001\\
Kwa-Dlangezwa, 3886, South Africa.\\
beeshama@unizulu.ac.za}

\maketitle

\begin{history}
\received{Day Month Year}
\revised{Day Month Year}
%\accepted{Day Month Year}
%\comby{(xxxxxxxxxx)}
\end{history}

\begin{abstract}
The general solution of the field equations in LRS Bianchi-I space-time with perfect fluid equation-of-state (EoS) is presented. The models filled with dust, vacuum energy, Zel'dovich matter and disordered radiation are studied in detail. A unified and systematic treatment of the solutions is presented, and some new solutions are found. The dust, stiff matter and disordered radiation models describe only a decelerated universe, whereas the vacuum energy model exhibits a transition from a decelerated to an accelerated phase.
\end{abstract}


\section{Introduction}

Several studies have been carried out on the Bianchi type I cosmological models, which  represent the simplest generalisation of the flat Friedmann-Lemaitre-Robertwson-Walker (FRW) models. Various methods have been employed to solve the field equations. In the  1960's, Thorne\cite{Thorne1967} presented some solutions of spatially homogeneous axisymmetic anisotropic open, semi-closed and Euclidean models with perfect fluid and magnetic fields and studied anisotropy and elements formation. Jacobs\cite{Jacobs1968} extended the work to the most general Euclidean Bianchi I models following an approach developed by Misner\cite{Misner1967,Misner1968}. Solutions of Einstein's equations for a fluid which exhibit local-rotational-symmetry (LRS) were presented by Stewart and Ellis\cite{StewartEllis1968}.\\
\indent In the 1980's, Hajj-Boutros\cite{Hajj-Boutros1984,Hajj-Boutros1985} introduced a technique to reduce the Einstein field equations to first-order Riccati equations in spherical symmetry. The authors applied this technique for generating exact solutions of LRS Bianchi type I models filled with a perfect fluid for which the classical barotropic equation-of-state (EoS) $p= (\gamma-1) \rho$ does not hold\cite{Hajj-Boutros1986cqg}. Further, the authors generated two new classes of LRS Bianchi type II models with stiff matter\cite{Hajj-Boutros1986jmp}. Hajj-Boutros and Sfeila\cite{Hajj-BoutrosSfeila1986grg} elaborated this new generation technique in the case of a static spherically-symmetric distribution of charged fluid satisfying a barotropic equation of state, i.e., $p= (\gamma-1) \rho$. In continuation of the series of their work, via a suitable scale transformation, they showed  that the condition of isotropy of pressure in a Bianchi I space-time filled with perfect fluid reduces to a linear second-order differential equation which can be used for generating many new LRS Bianchi I solutions\cite{Hajj-BoutrosSfeila1987ijtp}. Following their approach, Ram\cite{Ram1987ass,Ram1989grg,Ram1989ijtp}, and Singh and Ram\cite{SinghRam} also added some new classes of LRS Bianchi type I and type V$I_0$ perfect fluid models. In 1994, Mazumder\cite{Mazumder1994} showed that the field equations of the LRS Bianchi I space-time filled with a perfect fluid are solvable for arbitrary cosmic scale functions. He tried to generalise the solutions found in Refs.~\refcite{Hajj-BoutrosSfeila1987ijtp,Ram1989ijtp}. However, the main issue with generating schemes is that the matter energy density $\rho$ and pressure $p$ do not satisfy a barotropic equation of state, in general. Another weak point of these models is the consideration of known solutions to obtain exact solution of the field equations. In this paper, we discuss the general solution of the field equations in LRS Bianchi I space-time with a perfect fluid equation of state. \\
\indent The paper is organized as follows. In Sec. 2, we present the field equations and general solution with a perfect fluid equation of state in the framework of the LRS Bianchi I space-time. In subsections 2.1-2.5, we present the solutions for the dust, vacuum energy, Zel'dovich stiff matter and radiation models, respectively, and discuss their physical and cosmological significance. The summary of the results is accumulated in Sec. 3.

\section{The model and solution}

\indent The spatially homogenous and anisotropic LRS Bianchi I line-element is given by
\begin{equation}
ds^{2} =-dt^{2}+A^2dx^2+B^2(dy^{2}+dz^2),
\end{equation}
where $A$ and $B$ are the scale factors, and are functions of cosmic time $t$.\\
\indent The energy-momentum tensor for the perfect fluid is given as
 \begin{equation}
  T_{\mu\nu}=(\rho+p)u_\mu u_\nu+p g_{\mu\nu},
\end{equation}
\noindent where $\rho$ is the energy density and $p$ is the thermodynamical pressure of the fluid, $u_\mu$ is the four velocity of the fluid such that $u_\mu u^\mu=-1$ and in comoving coordinates $u^\mu=\delta_0^\mu$.\\
\indent The Einstein field equations are
\begin{equation}\label{efe}
R_{ij}-\frac{1}{2}g_{ij}R=T_{ij}\;,
\end{equation}
where we have taken $8\pi G=1=c$. The above field equations for the metric (1) and the energy-momentum tensor (2), yield the following independent equations
\begin{eqnarray}
  \left(\frac{\dot B}{B}\right)^2+2\frac{\dot A\dot B}{A B}&=&\rho,\\
  \left(\frac{\dot B}{B}\right)^2+2\frac{\ddot B}{ B}&=&-p,\\
\frac{\ddot A}{B}+\frac{\ddot B}{B}+\frac{\dot A \dot B}{AB}&=&-p.
\end{eqnarray}
From (5) and (6), Mazumder\cite{Mazumder1994} found the condition for isotropy of pressure as
\begin{equation}
(\dot B A -B \dot A)B=l,
\end{equation}
\noindent  where $l$ is a constant of integration. \\
\indent Equations (4)--(6) are three independent equations with four unknowns namely $A$, $B$, $\rho$ and $p$. Therefore, to find the exact solution to the field equations we require a supplementary constraint for the consistency of the system. One may assume any relation between any two arbitrary physical quantities or variables. We consider the perfect fluid equation of state (EoS) which is defined as
\begin{equation}
  p=\omega\rho,
\end{equation}
where $\omega$ is the EoS parameter. If the system is to be consistent with causality and mechanically stable, then $-1\leq\omega\leq1$\cite{Bondi1965}.\\
\indent Substituting (8) in (5) and eliminating $\rho$ from (4) and (5), we get
\begin{equation}
  \frac{\dot B}{B}+\omega\frac{\dot A}{A}=-\left(\frac{1+\omega}{2}\right)\frac{\ddot B}{ \dot B},
\end{equation}
which is a second order differential equation in $B$ and a first order differential equation in $A$. The variables are thus separated and the equation becomes integrable. We solve it for $A$ to obtain
\begin{equation}
  A^\omega=\frac{m\,B^{-\frac{1+\omega}{2}}}{\dot B},
\end{equation}
where $m$ is an integration constant which must be positive for an expanding universe. Equation (10) is the most general solution of the LRS Bianchi I model filled with a perfect fluid. However, this solution contains the derivative of $B$, therefore, is itself a first order linear differential equation in $B$. Equation (10) cannot be solved due to the presence of $A(t)$. Therefore, one may explicitly solve (7) and (10) by supplying  values to the EoS parameter $\omega$ for various forms of matter. The most common sources of the matter in the universe are non-relativistic matter (dust), ultra-relativistic matter (radiation), Zel'dovich matter (stiff matter) and vacuum energy (the cosmological constant). We shall determine the solution for $\omega=0$ (dust), $\omega=-1$ (vacuum energy), $\omega=1$ (stiff matter) in the forthcoming subsections and also study the influence of each of these matter sources in cosmological evolution. Let us define some cosmological parameters to study the cosmological evolution in LRS Bianchi I space-time.\\
\indent The average scale factor is defined as
\begin{equation}
  a=(AB^2)^{\frac{1}{3}}.
\end{equation}
\noindent The rates of the expansion along the  $x$, $y$, and $z -$axes are defined by
\begin{equation}
  H_x=\frac{\dot A}{A}, \;\;H_y=H_z=\frac{\dot B}{B},
\end{equation}
\noindent where a dot denotes the ordinary derivative with respect to cosmic time $t$. The average Hubble parameter (average expansion rate) $H$, which is the generalization of the Hubble parameter in the isotropic case, is given by
\begin{equation}
  H=\frac{1}{3}\left(\frac{\dot A}{A}+2\frac{\dot B}{B}\right).
\end{equation}
The streamlines of the motion of a cosmic fluid are characterized kinematically by their  expansion, $\theta$,  shear, $\sigma$, and rotation, $w$. Consider now a time-like congruence with a tangent vector $u^\mu$. Since any four dimensional quantity can be resolved into its space and time components by projecting it into the three dimensional space orthogonal to the time-like worldlines  by means of the operator $h_{\mu\nu}$,  then $u^\mu_{\;;\nu}$ may be decomposed as follows\cite{Urankar1970}
\begin{equation}
  u_{\mu;\nu}=w_{\mu\nu}+\sigma_{\mu\nu}+\frac{1}{3}\theta h_{\mu\nu}-\dot u_\mu u_\nu,
\end{equation}
where $w_{\mu\nu}$ is the tensor of rotation (vorticity), $\sigma_{\mu\nu}$ is the shear tensor,  $h_{\mu\nu} =g_{\mu\nu}-u_\mu u_\nu$ is the projection tensor and $\dot u^\mu=u^\mu_{\;;\nu}u^\nu$ is the acceleration vector.\\
\indent The expansion scalar, $\theta$, and the shear scalar, $\sigma$, are, respectively,  defined by
\begin{equation}
 \theta=u^\mu_{\;;\mu}=u^\mu_{,\mu}+\Gamma^\mu_{\;\mu\nu}u^\nu=\frac{\dot A}{A}+2\frac{\dot B}{B}=3H,
\end{equation}
\begin{equation}
\sigma^2=\frac{1}{2}\sigma_{\mu\nu}\sigma^{\mu\nu}=\frac{1}{3}\left(\frac{\dot A}{A}-\frac{\dot B}{B}\right)^2,
\end{equation}
\noindent where the shear tensor, $\sigma_{\mu\nu}$ is defined as\cite{GoennerKowalewski1989}
\begin{equation}
 \sigma_{\mu\nu} =u_{(\mu;\nu)}-\dot u_{(\mu} u_{\nu)}-\frac{1}{3}\theta h_{\mu\nu},
\end{equation}
where round brackets denote symmetrisation, e.g., $ u_{(\mu\nu)}=\frac{1}{2}(u_{\mu\nu}+u_{\nu\mu})$. For the metric (1), the acceleration $\dot u^\mu$ and vorticity $w_{\mu\nu}$ turn out to be zero.\\
\indent The shear tensor, $\sigma_{\mu\nu}$, determines the distortion arising in the fluid flow leaving the volume invariant\cite{CooleyHoogen1994}. The expansion rates can be different in the different directions, unlike the Robertson-Walker models where the expansion rates are the same in all directions. The directional components of the shear tensor are
\begin{equation}
\sigma_{\;1}^1=\frac{2}{3}\left(\frac{\dot A}{A}-\frac{\dot B}{B}\right),\;\sigma_{\;2}^2=\sigma_{\;3}^3=-\frac{1}{3}\left(\frac{\dot A}{A}-\frac{\dot B}{B}\right),\;\;\sigma_{\;4}^4=0\;\;\; \text{and}\;\;\; \sigma^\mu_{\;\nu}=0, \mu\neq\nu.
\end{equation}
The quantities we have defined here (expansion and shear) are called the kinematic quantities because they characterize the kinematic features of the fluid flow. For an expanding model $\theta>0$ and the shear decreases with time. The rate of work done by anisotropic stresses augments the shear dissipation. In the case of expanding models, it is found that the dynamical importance of matter increases while that of shear decreases in the course of evolution. In the spatially homogenous cosmological models in which the matter content of space-time is a perfect fluid and in which the fluid flow vector is not normal to the surfaces of homogeneity, the matter may move with non-zero expansion, shear and rotation. The only shear-free spatially homogenous perfect fluid universes models are Robertson-Walker models\cite{KingEllis1973}. \\
\indent Other than the above kinematical parameters one of the most important parameter for the present study is the deceleration parameter $q=-\frac{a\ddot a}{\dot a^2}$, which in terms of hubble parameter reads as
\begin{equation}
  q=-1-\frac{\dot H}{ H^2}.
\end{equation}
A positive deceleration parameter corresponds to a decelerated universe, whereas a negative one represents an
accelerating universe. In what follows we shall study the dust, vacuum energy and Zel'dovich stiff matter models.

\subsection{The dust model }

The dust model corresponds to $p=0$, i.e., $\omega=0$. Consequently, (10) reduces to
\begin{equation}
  \dot B=\frac{m}{\sqrt B},
\end{equation}
which on integration yields
\begin{equation}
   B(t)=\left(\frac{3mt}{2}\right)^{\frac{2}{3}}.
   \end{equation}
We have dropped out the integration constant so that the big-bang singularity occurs at $t=0$. We have already been considered $m>0$ for an expanding universe. In (21), the positivity of $m$ ensures the reality of the solution.\\
\indent Substituting (20) and (21) in  (7), we get
\begin{equation}
   A(t)=c_1 t^\frac{2}{3}+\left(\frac{2}{3m}\right)^\frac{4}{3}\frac{l}{t^\frac{1}{3}},
\end{equation}
where $c_1$ is an integration constant. We must have $l\geq0$ for an expanding universe. Since $A\to\infty$ as $t\to0$, therefore, the universe explodes with infinite rate of expansion in the direction of $A$. This is an example of what is called a cigar-type singularity, i.e., the expansion parameter tends to zero in 2 directions, but diverges in the last direction as $t\to 0$\cite{MacCullam1971}. This is due in this case to the presence of the inverse term $t^{-\frac{1}{3}}$ in (22). However, the expansion slows down as time passes and it speeds up once again as the first term in (19) starts dominating. From (20) and (22), we note that $A$ and $B$ are related by $A= \left(\frac{2}{3m}\right)^\frac{2}{3}B+\frac{l}{B^2}$. \\
\indent The solution in metric form can be written as
\begin{equation}
  ds^2=-dt^2+\left(c_1t^\frac{2}{3}+\frac{c_2}{t^\frac{1}{3}}\right)^2dx^2+c_3t^\frac{4}{3}(dy^2+dz^2),
\end{equation}
where $c_2=l\left(\frac{2}{3m}\right)^\frac{4}{3}$ and $c_3=\left(\frac{3m}{2}\right)^\frac{2}{3}$.\\
\indent The average scale factor is
\begin{equation}
  a=\left[{lt+ \left(\frac{3  m}{2}\right)^\frac{4}{3} t^2}\right]^\frac{1}{3},
\end{equation}
\noindent where we have taken the integration constant $c_1$ to be unity without any loss of generality. If $l=0$ then $a(t)\propto t^\frac{2}{3}$, i.e., the solution reduces to the Einstein-de Sitter solution for dust which has homogeneous and isotropic spatial sections. \\
\indent The average Hubble parameter gives
\begin{equation}
  H=\frac{2^\frac{4}{3} 3^\frac{2}{3} l+18 m^\frac{4}{3} t}{3\left(2^\frac{4}{3}  l +9 m^\frac{4}{3} t\right)t}.
\end{equation}
\noindent The energy density of dust matter becomes
\begin{equation}
  \rho=\left[\frac{l t}{2^\frac{2}{3} (3m)^\frac{4}{3}}+\frac{ t^2}{4} \right]^{-1},
\end{equation}
which remains always positive. It is to be noted that the energy density is infinite at $t=0$ which decreases with evolution of the universe and vanishes as $t\to\infty$.\\
\indent The deceleration parameter gives
 \begin{equation}
q=\frac{3 \left[ 2^\frac{8}{3} \sqrt[3]{3}\ l^2+ 3^\frac{5}{3} l (2m)^\frac{4}{3} t+27 m^\frac{8}{3} t^2\right]}
{2 \left(\sqrt[3]{2}\ 3^\frac{2}{3} l+9 m^\frac{4}{3} t\right)^2}.
\end{equation}
\begin{figure}[h]
  \centering
\includegraphics[width=8 cm]{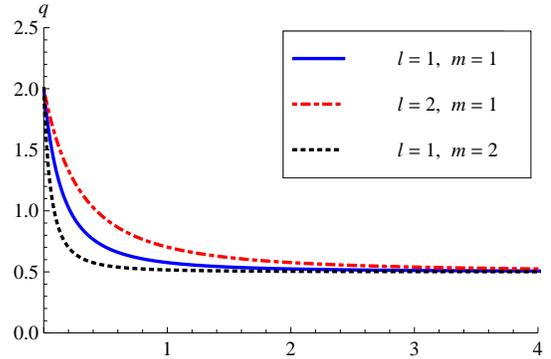}
  \caption{%
  $q(t)$ {\it \emph{\emph{versus}}} $t$ for different values of $l$ and $m$.}\label{}
\end{figure}
\noindent Figure 1 plots $q(t)$ \emph{versus} $t$, which shows that the deceleration parameter starts from $q=2$ at $t=0$ and approaches to $q\to0.5$ as $t\to\infty$. Since the deceleration parameter  always remains positive,  the dust model describes a decelerated phase of the universe. \\
\indent The shear scalar ($\sigma$) and the expansion scalar ($\theta$) have the expressions
\begin{equation}
  \theta=\frac{2^\frac{4}{3} 3^\frac{2}{3} l+18 m^\frac{4}{3} t}{\left(2^\frac{4}{3} 3^\frac{2}{3} l +9 m^\frac{4}{3} t\right)t},
\end{equation}
\begin{equation}
 \sigma= \frac{2^\frac{4}{3} 3^\frac{1}{6} l}{\left(2^\frac{4}{3}  3^\frac{2}{3} l +9 m^\frac{4}{3}t\right)t}.
\end{equation}
Consequently, the ratio of the expansion scalar to the shear scalar can be expressed as
\begin{equation}
  \frac{\sigma}{\theta}=\left(\frac{3\ 3^{5/6} m^{4/3} t}{\sqrt[3]{2} l}+\sqrt{3}\right)^{-1}.
\end{equation}
We have $\frac{\sigma}{\theta}=\frac{1}{\sqrt3}$ at $t=0$ and $\displaystyle\lim_{t\to\infty}\frac{\sigma}{\theta}=0$, therefore, the model is anisotropic at  early times but tends to isotopy for late times which is consistent with observations\cite{Netterfieldetal2002,Spergeletal2003,Bennettetal2013,Andersonetal2013,Adeetal2016}.\\
\indent The solution presented in here is a special case of the general Bianchi type I solution found by Robinson\cite{Robinson1961} for a universe containing only dust. Kompaneets and Chernov\cite{KompaneetsChernov1964} first obtained the axially symmetric solution in a different notation. Later on, the solutions were also obtained by Vajk and Eltgroth\cite{VajkEltgroth1970} by transforming independent variables and rediscovered by Iyer and Vishveshwara\cite{IyerVishveshwara1987}. The general solution has also been given in a different form by Jacobs\cite{Jacobs1968}. Similar solutions have also been obtained by Hajj-Boutros and Sfeila\cite{Hajj-BoutrosSfeila1987ijtp} applying a reverse way of generating technique to the Einstein-de Sitter metric. Ram\cite{Ram1989grg} obtained the above solution from the Kasner vacuum metric\footnote{The Kasner vacuum solution is given by $$ds^2=-dt^2+t^{2a}dx^2+t^{2b}(dy^2+dz^2)\;\;\text{where}\;\; a=-\frac{1}{3}\;\;\text{and}\;\;b=\frac{2}{3}.$$} by implementing the generating technique. In another paper, Ram\cite{Ram1989ijtp} obtained a similar solution starting from the dust-filled solution of Hajj-Boutros and Sfeila\cite{Hajj-BoutrosSfeila1987ijtp}.

\subsection{The vacuum energy model}

\indent The expansion of the universe is accelerating in the present epoch\cite{Riess1998,Perlmutteretal1999,Schmidtetal1998}. But the Einstein field equations lead to a decelerated expansion of the universe if the matter content is only ordinary baryonic matter. The accelerated expansion can be described by supplying some exotic component of the matter to the field equations. An unknown matter called dark energy (DE) is supposed to be responsible for the present accelerating universe. The past two decades have produced a flood of candidates for DE. However, a cosmological constant, $\Lambda$ is not only the simplest candidate for DE, which represents vacuum energy\cite{SahniStarobinsky2000,Padmanabhan2003,PeeblesRatra2003,Carroll2001}, but it also fits well with recent observations\cite{Adeetal2016,Komatsuetal2011}. \\
\indent Vacuum energy corresponds to $\omega=-1$, for which (10) gives
\begin{equation}
  A=\frac{\dot B}{m}.
\end{equation}
Consequently, (7) can be written as
\begin{equation}
  \ddot BB^2-\dot B^2 B=k,
\end{equation}
\noindent where $k=lm$. The above equation possesses two real solutions
\begin{eqnarray}
  B_1(t)&=&e^{-\sqrt{\beta} t} \left(\frac{e^{3 \sqrt{\beta} t}}{6\beta}- k\right)^\frac{2}{3},\\
  B_2(t)&=&e^{-\sqrt{\beta} t} \left( k e^{3 \sqrt{\beta} t}-\frac{1}{6 \beta}\right)^\frac{2}{3},
\end{eqnarray}
where $\beta$ is a positive integration constant. Substituting the above expressions in (31), we get
 \begin{eqnarray}
  A_1(t)&=&\frac{e^{-\sqrt{\beta} t} \left(e^{3 \sqrt{\beta} t} +6 \beta k\right)}{6^\frac{2}{3} \beta^\frac{1}{6} m \left({e^{3 \sqrt{\beta} t}-6 \beta k}\right)^\frac{1}{3}},\\
  A_2(t)&=&\frac{e^{-\sqrt{\beta} t} \left(6 \beta k e^{3 \sqrt{\beta} t} +1\right)}{6^\frac{2}{3} \beta^\frac{1}{6} m \left({6 \beta k e^{3 \sqrt{\beta} t}-1}\right)^\frac{1}{3}}.
\end{eqnarray}
\noindent The average scale factors for both solutions become, respectively,
\begin{eqnarray}
  a_1(t)&=&\left[\frac{e^{-3 \sqrt{\beta} t} \left(e^{6 \sqrt{\beta} t}-36 \beta^2 k^2 \right)}{36 \beta^\frac{3}{2} m}\right]^\frac{1}{3},\\
  a_2(t)&=&\left[\frac{e^{-3 \sqrt{\beta} t} \left(36 \beta^2 k^2  e^{6 \sqrt{\beta} t}-1\right)}{36 \beta^\frac{3}{2} m}\right]^\frac{1}{3}.
\end{eqnarray}
The average Hubble parameters give
\begin{eqnarray}
  H_1&=&\left(\frac{1}{\sqrt{\beta }}-\frac{72 \beta ^{3/2} k^2}{36 \beta ^2 k^2+e^{6 \sqrt{\beta } t}}\right)^{-1},\\
H_2&=&\sqrt\beta_1\left(1-\frac{2}{36 \beta ^2 k^2 e^{6 \sqrt{\beta } t}+1}\right)^{-1}.
\end{eqnarray}
The energy density and pressure of vacuum energy are constants, i.e., $\rho_1=3\beta=-p$. Thus the integration constant $\beta$ stands for a cosmological constant which represents the vacuum energy in the present model. \\
\indent Let us first discuss the model for the solution of $A_1$ and $B_1$. The deceleration parameter takes the form
 \begin{equation}
q_1=-1+\frac{432 \beta ^2 k^2 e^{6 \sqrt{\beta } t}}{\left(36 \beta ^2 k^2+e^{6 \sqrt{\beta } t}\right)^2}.
\end{equation}
\begin{figure}[h]
  \centering
\includegraphics[width=8 cm]{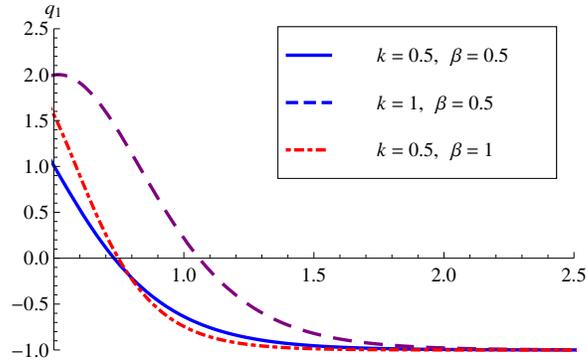}
  \caption{%
  $q_1$ {\it \emph{\emph{versus}}} $t$ for different values of $k$ and $\beta$.}\label{}
\end{figure}
Figure 2 plots $q_1(t)$ \emph{versus} $t$ which shows the transition from a decelerated to an accelerated universe which is consistent with many recent observations\cite{Riessetal2001,Amendola2003,Riessetal2004,Riessetal2007,Lietal2011,Giostrietal2012}.\\
\indent The expansion scalar and the shear scalar have  expressions
 \begin{equation}
\theta_1=3\left(\frac{1}{\sqrt{\beta }}-\frac{72 \beta ^\frac{3}{2} k^2}{36 \beta ^2 k^2+e^{6 \sqrt{\beta } t}}\right)^{-1},
\end{equation}
\begin{equation}
  \sigma_1=\frac{12\sqrt3\beta^\frac{3}{2} k e^{3 \sqrt{\beta} t}}{e^{6 \sqrt{\beta} t}-36 \beta^2 k^2}.
\end{equation}
The ratio of shear scalar to expansion scalar can be written as
\begin{equation}
  \frac{\sigma_1}{\theta_1}= \frac{4 \sqrt{3} \beta  k e^{3 \sqrt{\beta } t}}{36 \beta ^2 k^2+e^{6 \sqrt{\beta } t}}.
\end{equation}
\begin{figure}[h]
  \centering
\includegraphics[width=8 cm]{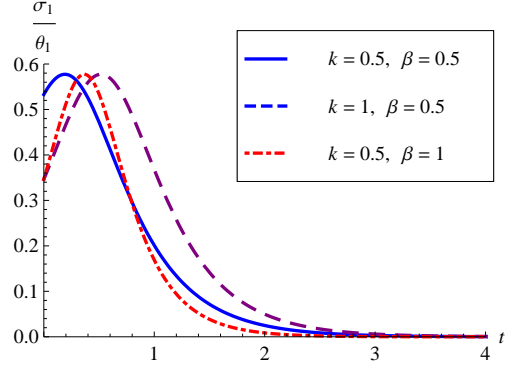}
  \caption{%
  $\frac{\sigma_1}{\theta_1}$ {\it \emph{\emph{versus}}} $t$ for different values of $k$ and $\beta$.}\label{}
\end{figure}
\noindent Figure 3 plots $\frac{\sigma_1}{\theta_1}$ \emph{versus} $t$ which shows that the universe was anisotropic at early times but becomes isotropic at late times.\\
\indent Now for the second solution corresponding to $H_2$, the deceleration parameter takes the form
 \begin{equation}
q_2=-1+\frac{432 \beta ^2 k^2 e^{6 \sqrt{\beta } t}}{\left(1+36 \beta ^2 k^2 e^{6 \sqrt{\beta } t}\right)^2}.
\end{equation}
\begin{figure}[h]
  \centering
\includegraphics[width=8 cm]{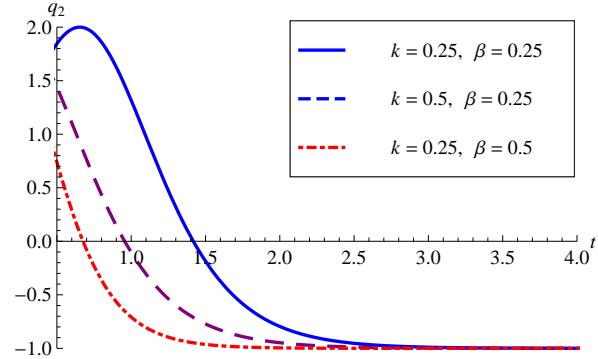}
  \caption{%
  $q_2$ {\it \emph{\emph{versus}}} $t$ for different values of $k$ and $\beta$.}\label{}
\end{figure}
Figure 4 plots $q_2$ \emph{versus} $t$, which also describes the transition from a decelerated to an accelerated phase of the universe.\\
\indent The expressions for expansion scalar and shear scalar are
\begin{eqnarray}
  \theta_2&=&3 \sqrt{\beta}\left(1-\frac{2}{1+36 \beta^2 k^2  e^{6 \sqrt{\beta} t}}\right)^{-1},\\
  \sigma_2&=&\frac{12 \sqrt{3} \beta ^\frac{3}{2} k e^{3 \sqrt{\beta } t}}{36 \beta ^2 k^2 e^{6 \sqrt{\beta } t}-1},
\end{eqnarray}
respectively. Consequently
\begin{equation}
  \frac{\sigma_2}{\theta_2}=\frac{4 \sqrt{3} \beta  k e^{3 \sqrt{\beta } t}}{1+36 \beta ^2 k^2 e^{6 \sqrt{\beta } t}}.
\end{equation}
One may observe that the behavior of $\frac{\sigma_2}{\theta_2}$ is similar to that shown in Fig. 3. Therefore, this case also shows the anisotropic behavior of the model at early times which becomes isotropic at late times. Thus, the characteristics of the models with both solutions of $a_1$ and $a_2$ are similar.\\
\indent It is to be noted that Iyer and Vishveshwara\cite{IyerVishveshwara1987} also found solution for constant vacuum energy density. The present solutions are different from of those obtained in Ref.~ \refcite{IyerVishveshwara1987}, since we have not assumed a constant vacuum energy density. Rather, it is the natural outcome of our procedure. To the best of our knowledge the solutions obtained here are new.

\subsection{Zel'dovich stiff matter model}

The Zel'dovich stiff matter corresponds to $\omega=1$\cite{Zel'dovich}. In this case  (10) reduces to
\begin{equation}
A= \frac{ m}{\dot BB}.
\end{equation}
Consequently, (7) can be written as
\begin{equation}
m\left(\frac{\ddot BB}{\dot B^2}+2\right)-n=0,
\end{equation}
where we have taken $l=-n$ ($n>0$) for reality of the solution. The above equation yields
\begin{equation}
B(t)=\left\{
  \begin{array}{ll}
     \beta \left[ (n+3 m)t\right]^{\frac{m}{n+3 m}}, & \hbox{$n\neq3m$;} \\
    \beta e^{\alpha  t}, & \hbox{$n=3m$,}
  \end{array}
\right.
\end{equation}
Here $\alpha$ and $\beta$ are constants of integration and one integration constant in case of $n\neq3m$, has been taken zero so that the big-bang singularity occurs at $t=0$. We must have $\beta>0$ for an expanding universe.\\
\indent Substituting the values of $B(t)$  in (49), we get
\begin{equation}
A(t)=\left\{
  \begin{array}{ll}
     \frac{1}{\beta^2}\left[ (n+3 m)t\right]^{\frac{n+ m}{n+3 m}}, & \hbox{$n\neq3m$;} \\
    \frac{m e^{-2 \alpha  t}}{\alpha  \beta ^2}, & \hbox{$n=3m$.}
  \end{array}
\right.
\end{equation}
\noindent The directional scale factors are related by $A=\frac{B}{\beta}\left[(n+3m)t\right]^\frac{n}{n+3m}$ for $n\neq3m$ and $A=\frac{mBe^{-3\alpha t}}{\alpha\beta^3}$ for $n=3m$. For $n\neq3m$, $A=0=B$ at $t=0$, which shows a point type singularity, whereas for $n=3m$, $B=\beta$ and $A=\frac{m}{\alpha\beta^2}$ at $t=0$, which is a singularity-free model. Let us express both solutions in metric form:
\begin{eqnarray}
  ds^2&=&-dt^2+c_4t^\frac{2(n+m)}{n+3m}dx^2+c_5t^\frac{2m}{n+3m}\left(dy^2+dz^2\right); \;\;n\neq3m,\\
  ds^2&=&-dt^2+c_6e^{-4\alpha t}dx^2+c_7e^{2\alpha t}\left(dy^2+dz^2\right);\;\;n=3m,
\end{eqnarray}
\noindent where $c_4=\frac{(n+3m)^\frac{2(n+m)}{n+3m}}{\beta^4}$, $c_5=\beta^2(n+3m)^\frac{2m}{n+3m}$, $c_6=\frac{m^2}{\alpha^2\beta^4}$ and $c_7=\beta^2$. These metrics represent the most general solutions of the stiff matter model in the LRS Bianchi I spacetime model, which are completely different from those  obtained by generating methods in previous works\cite{Hajj-BoutrosSfeila1987ijtp,Ram1989grg}. If we choose $\frac{m}{n+3m}=k$ then one of the solutions (53) can be represented by a one-parameter family of solutions to Einstein's equation with a perfect stiff-matter fluid first obtained by Jacobs\cite{Jacobs1968}, i.e.,
\begin{equation}
  ds^2=-dt^2+c_4t^{2(1-2k)}dx^2+c_5t^{2k}\left(dy^2+dz^2\right).
\end{equation}
\noindent It is to be noted that the above solution is different from the general LRS Kasner stiff-matter metric\footnote{The general LRS Kasner stiff-matter metric is of the form $$ds^2=-dt^2+t^{2a}dx^2+t^{2b}(dy^2+dz^2)\;\;\text{where}\;\; a+2b=1.$$}. Jacobs also discussed the nature of the singularity in detail for this solution. Vajk and Eltgroth\cite{VajkEltgroth1970} found some general solutions for rational values between $-1<\omega<1$ and a particular solution for stiff matter with different parameterizations. Later on, Iyer and Vishveshwara\cite{IyerVishveshwara1987} rediscovered stiff matter solutions identical to (55) in searching for exact solutions of the Einstein equations in which the Dirac equation separates. The present solutions are different from the solutions of Hajj-Boutros and Sfeila\cite{Hajj-BoutrosSfeila1987ijtp} obtained by applying a  generating technique to a flat FRW metric with unity expansion rate. The solutions obtained by Ram\cite{Ram1989grg} by implementing generating technique to LRS Kasner stiff-matter metric do not satisfy classical EoS of perfect fluid.\\
\indent In particular, if $m=-\frac{2n}{3}$ then the solution is given in (53) reduces to
\begin{equation}
   ds^2=-dt^2+c_4t^{-\frac{2}{3}}dx^2+c_5t^\frac{4}{3}\left(dy^2+dz^2\right).
\end{equation}
The above metric is identical to the solution of Singh and Ram\cite{SinghRam}, which also does not satisfy a perfect fluid equation of state due to following the solutions generating method. \\
\indent The average scale factor is given as
\begin{equation}
a(t)=\left\{
  \begin{array}{ll}
     \left[(n+3 m)t\right]^\frac{1}{3}, & \hbox{$n\neq3m$;} \\
    \left(\frac{m}{\alpha }\right)^\frac{1}{3}, & \hbox{$n=3m$.}
  \end{array}
\right.
\end{equation}
The scale factor for $n\neq3m$ describes a power-law expansion of the universe, whereas the scale factor for $n=3m$ corresponds to a static universe. However, only the volume remains constant for the static universe and we can see that the shape of the universe changes exponentially in the spatial directions of $A$ and $B$. \\
\indent The average Hubble parameter is given by
\begin{equation}
H=\left\{
  \begin{array}{ll}
     \frac{1}{3t}, & \hbox{$n\neq3m$;} \\
    0, & \hbox{$n=3m$.}
  \end{array}
\right.
\end{equation}
\noindent The deceleration parameter also has the constant values, $q=2$ for $n\neq3m$ and $q=0$ for $n=3m$. Hence, the Zel'dovich stiff matter model describes a decelerating universe for $n\neq3m$ and a marginal inflationary cosmology for $n=3m$. \\
\indent For stiff matter, the energy density and pressure are equal. In the present model they become
\begin{equation}
\rho(=p)=\left\{
  \begin{array}{ll}
     \frac{m (2 n+3 m)}{t^2 (n+3 m)^2}=\frac{k(2-3k)}{t^2}, & \hbox{$n\neq3m$;} \\
    -3\alpha^2, & \hbox{$n=3m$.}
  \end{array}
\right.
\end{equation}
For $n\neq3m$, the energy density (or pressure) decreases with the evolution of the universe and vanishes as $t\to\infty$. The energy density is negative for $n=3m$, which does not represent a realistic model of the universe. \\
\indent The expansion and shear scalars are  given by, respectively
\begin{eqnarray}
\theta&=& \left\{
  \begin{array}{ll}
     \frac{1}{ t}, & \hbox{$n\neq3m$;} \\
    0, & \hbox{$n=3m$.}
  \end{array}
\right.,\\
\sigma&=&\left\{
  \begin{array}{ll}
     \frac{n}{\sqrt{3}  (n+3 m)t}, & \hbox{$n\neq3m$;} \\
    \sqrt3\alpha, & \hbox{$n=3m$.}
  \end{array}
\right.
\end{eqnarray}
The ratio of the shear scalar to the expansion scalar for $n\neq3m$ has the constant value $\frac{\sigma}{\theta}=\frac{n}{\sqrt3(n+3m)}$, which shows that the stiff matter model remains always anisotropic for all finite values of $n\neq3m$. However, the model becomes isotropic  in the case of $n\neq3m$ when $n\to0$ or $m\to\infty$. There is no expansion of the universe for $n=3m$, but it has finite shear $\sqrt{3}\alpha$.

\subsection{Disordered radiation model}

Klein\cite{Klein1947} and Teixeira et al.\cite{Teixeiraetal1977} investigated a source free disordered distribution of electromagnetic radiation. The EoS of disordered radiation is $p=3\rho$. In this case (10) reduces to
\begin{equation}
A^3= \frac{ m}{\dot BB^2}.
\end{equation}
Consequently, (7) can be written as
\begin{equation}
B^3 \left(\frac{m}{\dot BB^2 }\right)^\frac{4}{3} \left(\ddot BB +5 \dot B^2\right)=3lm.
\end{equation}
\noindent The only real solution which above equation possesses is
\begin{equation}
B(t)=l \left({\frac{2 t^2}{3 m}}\right)^\frac{1}{3},
\end{equation}
where both integration constants have been taken zero, one is for the choice of the big-bang singularity at $t=0$ and another for the reality of the solution. For an expanding universe we have must have $l>0$ .\\
\indent Substituting (64) in (62), we get
\begin{equation}
A(t)=\left(\frac{3 m}{2}\right)^\frac{2}{3} \frac{1}{lt^\frac{1}{3}}.
\end{equation}
\noindent Form (64) and (65), the directional scale factors are related by $B=\frac{2 l^2 A\,t}{3m}$. The solution in metric form can be written as
\begin{equation}
   ds^2=-dt^2+c_8t^{\frac{2}{3}}dx^2+c_9t^\frac{4}{3}\left(dy^2+dz^2\right),
\end{equation}
\noindent where $c_8=l\left(\frac{3 m}{2}\right)^\frac{2}{3}$ and $c_9=l \left({\frac{2 }{3 m}}\right)^\frac{1}{3}$. As far as we are aware, the above metric adds a new class of solutions to the  LRS Bianchi I model.\\
\indent The average scale factor is given by
\begin{equation}
a(t)=\left(l t\right)^\frac{1}{3}.
\end{equation}
\noindent The average Hubble parameter, $H=\frac{1}{3t}$, deceleration parameter, $q=2$, and expansion scalar, $\frac{1}{t}$, are similar to the solution of the Zel'dovich model for $n\neq3m$. Therefore, the disordered radiation model describes a decelerating universe. The energy density and pressure vanish, i.e., $\rho=0=p$, which shows that the disordered radiation is source free.\\
\indent The shear scalar is given by
\begin{equation}
\sigma=\frac{1}{\sqrt{3}\;  t}.
\end{equation}
The ratio of the shear scalar to the expansion scalar has a constant value $\frac{\sigma}{\theta}=\frac{1}{\sqrt3}$, which shows that the universe filled with disordered radiation remains anisotropic.\\
\indent This solution is  different from that of Hajj-Boutros and Sfeila\cite{Hajj-BoutrosSfeila1987ijtp} obtained from the Tolman metric for disordered radiation. The classical EoS does not hold for their solution, and in particular leads to  stiff matter.

\subsection{The radiation model}

\indent Ultra-relativistic radiation corresponds to $\omega=\frac{1}{3}$, for which (10) gives
\begin{equation}
  A=\frac{m^3}{\dot B^3 B^2}.
\end{equation}
Consequently, (7) can be written as
\begin{equation}
  3 m^3 \left(B \ddot B+\dot B^2\right)=l\dot B^4B ,
\end{equation}
\noindent The above equation possesses two real solutions
{\tiny\begin{eqnarray}
  B_1(t)&=&\text{InverseFunction}\left[-\frac{\sqrt{3} m^{3/2} t \left(2 l^2+9 l m^3 t+9 m^6 t^2\right)-2 l^2 \sqrt{t} \sqrt{2 l+3 m^3 t} \log \left(\sqrt{3} m^{3/2} \sqrt{2 l+3 m^3 t}+3 m^3 \sqrt{t}\right)}{18 m^6 \sqrt{t \left(2 l+3 m^3 t\right)}}\right],\\
  B_2(t)&=&\text{InverseFunction}\left[\frac{\sqrt{3} m^{3/2} t \left(l+3 m^3 t\right) \left(2 l+3 m^3 t\right)-2 l^2 \sqrt{t} \sqrt{2 l+3 m^3 t} \log \left(m^{3/2} \sqrt{6 l+9 m^3 t}+3 m^3 \sqrt{t}\right)}{18 m^6 \sqrt{t \left(2 l+3 m^3 t\right)}}\right],
\end{eqnarray}}
\noindent where one integration constant has been taken zero and another  unity. Since the above expressions involve complicated inverse functions,  it is not possible to give a simple  physical interpretation in this case. However, one may also write the scale factors $A_1$ and $A_2$ for the above expressions which would be more complicated expressions of inverse functions. It is to be noted that Iyer and Vishveshwara\cite{IyerVishveshwara1987} have  found the solution for radiation.

\section{Conclusion}

\noindent In this paper, we have presented the general solution of the field equations in LRS Bianchi-I space-time  with perfect fluid equation of state. In different cases of particular interest, we have studied dust, vacuum energy, Zel'dovich stiff matter and disordered radiation models. Though most of these solutions were known earlier, we present a unified and systematic treatment by solving the field equations in a straight forward manner. However, as far we know, the vacuum energy and disordered radiation solutions are new. It has been found that the dust, Zel'dovich stiff matter and disordered radiation models describe only decelerated universes, whereas the vacuum energy model exhibits a transition from a decelerated to an accelerated universe.\\
\indent It is well known that the anisotropic models may represent the cosmos during its early stages of  evolution. But the investigation of the vacuum energy model shows that the anisotropic models can also successfully describe a sudden change from deceleration to acceleration. The models describe anisotropic behavior at  early times and becomes isotropic at late times, except in the disordered radiation model and in a particular case of the Zel'dovich model. The disordered radiation model remains anisotropic throughout the evolution of the universe. \\
\indent The straight forward procedure used to solve the field equations is much more apealing. We hope that this will make it useful in future applications of anisotropic cosmological models. We shall explore  more solutions in other Bianchi space-time models in our future work.

\section*{Acknowledgement}

The authors are thankful to the reviewer for his constructive comments and suggestions to improve the quality of the manuscript. Vijay Singh expresses his sincere thanks to the University of Zululand, South Africa, for providing a postdoctoral fellowship. Vijay Singh extends his sincere thanks to Inter University Center for Astronomy and Astrophysics (IUCAA), Pune, India for providing hospitality and other necessary facilities where the part of the work has been completed.

\end{document}